\documentclass{epl}
\title{Stretching  Polymers in Poor  and  Bad Solvents: Pullout Peaks and an Unraveling
Transition }
\author{I. R. Cooke \inst{1} \and D.R.M. Williams\inst{1}}
\institute{
  \inst{1} Research School of Physical Sciences, Institute of Advanced Studies, Australian
National University, ACT 0200, Australia\\
}

\begin{document}

\maketitle

\begin{abstract}
Using analytic theory, numerical calculation and Langevin dynamics
simulation we demonstrate the existence of a first order unraveling
transition in the stretching of a polymer chain in a poor solvent.
The chain suddenly unravels from a ``tadpole" or ``ball and chain"
configuration, to one where the ball shrinks to zero size. In the
force curve this appears as a discontinuous drop in the force. This
transition occurs under the conditions of most relevance for atomic
force microscope experiments, where the extension is the independent
variable.  Our simulations show marked hysteresis as well as many
local peaks associated with the pullout of small portions of the
chain.
\end{abstract}

\begin{figure}
\onefigure[]{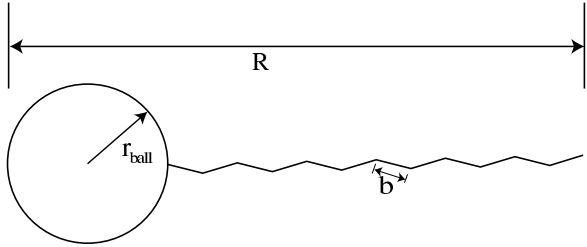}
\caption{ The ball and chain model used for analytic and numerical
calculations. $R$ is the total stretching distance, $r_{ball}$ is the
radius of the ball , $b$ is the link length.  In practice one often
finds two tails rather than one, but this does not affect the free
energy calculations.
\label{model}}
\end{figure}

%\section{}
The deformation of single polymer chains is one of the most
fundamental parts of polymer physics. One might expect therefore that
almost nothing remains to be discovered. This however is not the
case. Indeed, over the past decade several novel kinds of behaviour
have been uncovered for single chains subjected to simple extension or
compression or to hydrodynamic flows \cite{halperin_zhulina:1991,
brochardwyart:1993, brochardwyart:1994, subramanian:1995,
williams:1993, sevick_williams:1999}. Much of this work has been
inspired by experimental developments in imaging and deforming single
polymer chains, by atomic force microscopy, optical tweezers and
fluorescence microscopy
\cite{senden:1998,wuite:2000,haupt_senden_sevick:2002}.  Among the
most novel of these has been the transition undergone by a polymer in
a poor solvent when subjected to a stretching force, discovered by
Halperin and Zhulina a decade ago \cite{halperin_zhulina:1991}.  In
this transition, with force as the independent variable (FIV), a
collapsed globule first deforms from a sphere to an ellipsoid and then
at a critical force it totally unwinds. With extension as the
independent variable (EIV) the polymer undergoes a transition from a
globule to a tadpole or ball and chain configuration shown in figure
\ref{model}. As noted in \cite{buguin_brochard-wyart:1996} this is a
manifestation of the Rayleigh-Plateau instability
\cite{plateau:1873,rayleigh:1879,sevick_williams:1999} for this
system: a liquid cylinder is unstable to undulations and to the
formation of droplets.

This kind of system has been much-studied by computer simulation
\cite{lai:1996,lai:1998,maurice_matthai:1999}, and more recently by
experiment \cite{haupt_senden_sevick:2002}. The concentration has been
either on FIV conditions, or on systems very close to the $\theta$
temperature, where solvent effects are weak. However, it is EIV
conditions that are most relevant in experiments, and as we shall
show, if the system is quenched well below the $\theta$ temperature
some novel behaviour results. Under EIV conditions we show there are
in fact two transitions. The first one (from globule to tadpole) was
already analysed at length by previous authors
\cite{halperin_zhulina:1991}.  There is however a second, even more
dramatic transition, at higher extension, from a tadpole to a headless
tadpole. This we call the unraveling transition. We will show that
this transition is sharp (first order), and occurs when the ball or
tadpole head is still large.  A similar transition to this unraveling
transition has previously been predicted for polyelectrolytes in poor
solvent from a detailed analytic theory
\cite{tamashiro_schiessel:2000, vilgis_etal:2000}.  In that work,
discontinuous jumps in the force were observed as a result of the
unravelling of ``pearls'' in a ``pearl necklace'' model of a
polyelectrolyte. Here we show using a very simple model and computer
simulation that an analogous transition exists for neutral polymers
and that charges are totally unnecessary for this to occur.  This is in
marked contrast to earlier studies on neutral polymers which suggested
that as the extension was increased, the size of the ball decreased
gradually until it reached the thermal blob size
\cite{kreitmeier:2000, halperin_zhulina:1991}, which for poor solvent
is close to zero.  In this paper we show that this interpretation
overlooks some interesting physics.

There is an energy barrier between the states, making the transition
first order, and as a consequence the system will show very strong
hysteresis. Furthermore, in practice we show that the behaviour under
increasing extension is very different from that under relaxation. In
fact, under extension we see a very jagged force curve which
corresponds to monomers becoming stuck within the ball and needing
significant force to remove them. During relaxation, the monomers
accrete to the ball and the force curve is much smoother. We utilise
in turn a numerical minimisation of the free energy, a simplified
analytical calculation of the transition and then a detailed Langevin
computer simulation of the system. It is this latter piece of work
that in practice will be most like the experiments.  

Our model for a stretched polymer chain is summarised in figure
\ref{model}.  This simple geometrical model is based on theoretical
predictions \cite{halperin_zhulina:1991} and simulations
\cite{kreitmeier:2000,maurice_matthai:1999} which observe a ball and
chain geometry at intermediate to strong extension.  It is precisely
this intermediate to strong extension regime which we are interested
in. The regime at small extensions has been thoroughly studied
previously \cite{lai:1998,halperin_zhulina:1991}, and we do not
consider it in detail here.  Furthermore we consider the regime
ignored by previous authors, well below the theta temperature, where
all solvent is excluded from the ball.

Monomers are represented by spheres with radius $a$.  For a chain
stretched a distance $R$, with $N$ total monomers and $n$ monomers in
the tail the free energy is
\begin{equation}
F = g kT (N-n)^\frac{2}{3} + g^{\prime} kT n +
F_{stretch}(R-2r_{ball}, n)
\end{equation}
Here we have set $g \equiv 4 \pi \gamma a^2/kT$ where $\gamma$ is the
polymer-solvent surface tension. Hence $g$ is a dimensionless measure
of this surface tension and $g^{\prime}$ measures the energy penalty
for exposing a monomer in the tail to the solvent. In principle both
$g$ and $g^\prime$ arise from the same short-range dispersion forces,
and in practice one expects them both to be of the same order of
magnitude with $g^\prime$ being somewhat less than $g$. The first term
in the free energy represents the ball surface energy for a ball of
radius $r_{ball} = a (N-n)^{1/3}$. The second term accounts for the
solvent interaction with monomers in the stretched portion.  The
remaining term $F_{stretch}$ comes from the elastic stretching energy
required to stretch $n$ monomers a distance $R -2 r_{ball}$. Many
idealised models exist for $F_{stretch}$, and all would give the
transition we are discussing. Here we use the freely-jointed chain
model \cite{flory:1953} with link size $b$ so that
$F_{stretch}=kTb^{-1}\int_0^{R-2r_{ball}}dx\ell^*(x/(nb))$ where
$\ell^*$ is the inverse Langevin function.  It is useful to define a
new variable $\epsilon\equiv \frac{n}{N}$ , which is the fraction of
monomers in the tail

With the stretching distance $R$ as the independent variable there is
only one variable, $\epsilon$, to minimise the free energy over.
Provided the stretching distance $R$ is not too large there is usually
a minimum in $F$ at some finite value of
$\epsilon=\epsilon_{bc}(R)$. This minimum corresponds to the ordinary
ball-and-chain configuration. As the stretching distance is increased
a second minimum appears, at $\epsilon$ very close to $1$. This second
minimum is itself unphysical, since it corresponds to less than one
monomer in the ball. However, it is informing us that we should
examine the free energy of the state with no monomers in the ball,
i.e. $\epsilon=0$. We then need to compare $F(\epsilon_{bc})$ with
$F(\epsilon=0)$. When these two are equal, at $R=R_u$, a transition
takes place from a ball-and-chain to an unraveled chain.  As we shall
see, this unraveling occurs when the ball is still large.

\begin{figure}
\onefigure[]{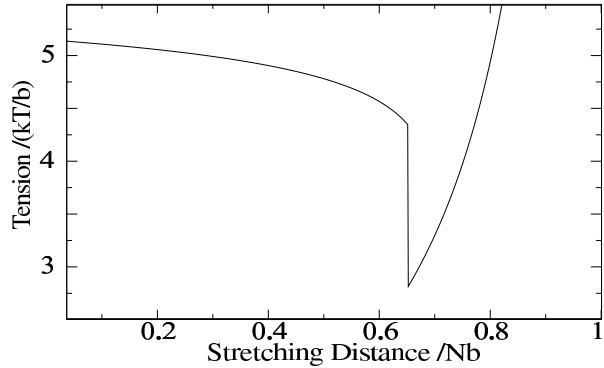}
\caption{ Numerical calculation of the chain tension vs stretching
distance showing the unraveling transition.  The following parameters
were used, $N=1000$, $g = 12$, $g^{\prime} = 3.6$, $b=1$.
\label{fvsr}}
\end{figure}

The most readily accessible quantity in experiments is the tension in
the chain $f = \partial F/\partial R$.  Figure \ref{fvsr} shows $f$
evaluated numerically as $R$ is varied towards its fully stretched
value of $bN$.  The unraveling transition mentioned above is clearly
evident at as a discontinuous change in $f$, that separates two
distinct stretching behaviours. In the region where $R < R_{u}$, the
tension decreases slightly and gradually with the extension.  As the
ball becomes smaller it becomes easier to remove monomers from it.  At
$R=R_{u}$ the tension drops dramatically as many new monomers are
suddenly added to the tail.  For $R> R_{u}$ the tension increases
rapidly as the fully pulled out chain is stretched towards its limit
$R = Nb$. The reader should note that this transition can only be seen
under EIV conditions. Using force as the independent variable would
lead to an instability in the region where the extension increases
with decreasing force.

Having demonstrated the existence of an unfolding transition at
$R_{u}$ numerically, we now present some analytic calculations.  These
are certainly not exact, but they do help to clarify some of the
physics, and yield a simple relation for the size of the ball at the
transition. As an approximation to the stretching energy we assume a
Gaussian chain, i.e. the first term in a series expansion of the
freely-jointed chain, or indeed of any other chain model. We also
assume that the radius of the ball can be neglected at the transition
point when calculating the stretching energy i.e. $R - 2 r_{ball}
\approx R$. This leads to a free energy:

\begin{equation}
F/kT = g N^{\frac{2}{3}} (1- \epsilon)^{\frac{2}{3}} + g^{\prime} N
\epsilon + \frac{3}{2}
\frac{R^2}{\epsilon N b^2}
\label{sfe}
\end{equation}

At the transition point the derivative of the first term is small
compared to the other two, so this term can be neglected in solving
$\frac{\partial F}{\partial \epsilon}=0$. Physically this means that
near the transition the chain conformation is a balance between the
stretching force on the tail and the surface tension of the tail, with
the ball playing little part.  At the transition we find
$\epsilon_{bc} = \sqrt{3/2} R/ \sqrt{g^{\prime}
Nb}$. Substituting this into (\ref{sfe}) (including now the ball term,
which is important, although its derivative is not), and equating to
the free energy of the unraveled chain yields a transition at a
critical distance of $R_{u} = \sqrt{2/3} b
{g^{\prime}}^{-\frac{1}{4}}N^{\frac{3}{4}} (
{g^{\prime}}^{\frac{3}{4}} N^{\frac{1}{4}} - g^{\frac{3}{4}} )$ with
the number of monomers in the ball at the transition being $N-n =
(1-\epsilon_{u})N = (\frac{g}{g^{\prime}})^\frac{3}{4}
N^{\frac{3}{4}}$

We can also show that at the transition the ratio $R_{u}$ to the fully
stretched length of the tail $ n b$ is $\sqrt{2/3} g^{\prime}
\approx 0.8 \sqrt{g^{\prime}}$ so that provided $g^{\prime}$ is close
to unity the Gaussian approximation for the stretching works
well. Indeed, a direct comparison with the numerical minimisation of
the full free energy shows this.  However, for higher surface tensions
our analytic results break down and we need to include higher order
terms in the free energy. While it is possible to make some analytic
progress doing this, the utility of such a calculation is
questionable, since the answer depends in detail on the model for the
stretching energy.

The most important point to note about our analytic results is that at
the transition the number of monomers in the ball is of order
$N^{\frac{3}{4}}$ and is thus large. The ball is thus a well-defined
object at the transition point, consisting of many monomers, and
indeed if the tail was removed the ball would itself be a stable
entity.

\begin{figure}
%\onefigure[]{fig3.eps}
\caption{ Simulation results for an individual run with a stretching
speed of $0.0001 b/t$ for $90$ monomers . The plot shows the  chain tension
(solid lines) and the number of monomers remaining in the ball (dashed
lines) for both pullout and relaxation.  In both cases the upper of
the two lines represents pullout and the lower line relaxation (refer
to figure \ref{simav} for clarification).  Alphabetic labels represent
precise points for which corresponding graphics of the chain
configuration are shown.  The unraveling and re raveling transitions
are most clear in the data for the number of monomers remaining in the
ball. ( See the attached figure 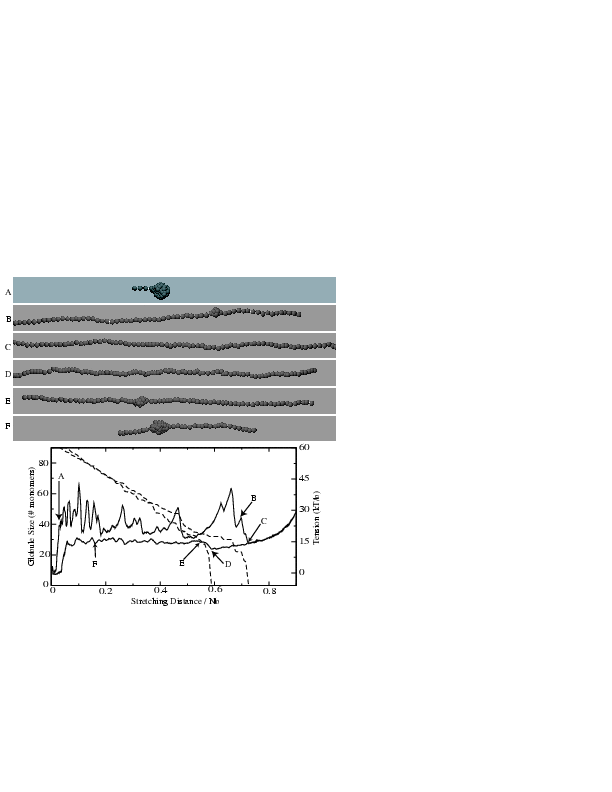 )
\label{fvsrsim}}
\end{figure}

All of this discussion assumes that the system is always in
equilibrium and the free energy is thus minimised globally. In most
practical experiments there will be a large energy barrier between the
ball and chain and unraveled states. This energy barrier leads
naturally to hysteresis if one performs an experiment where the
extension is gradually increased and then gradually reduced. In order
to test this, and more generally to test the existence of the
transition, we have performed free-draining Langevin dynamics
simulations (figure \ref{fvsrsim}) of stretching and relaxation for
polymers of different sizes.  Solvent monomer interactions were
accounted for by imposing a Lennard-Jones potential between monomers
separated by a distance $r$, which in our dimensionless units is,
$V_{LJ} = 2( r^{-12}- r^{-6})$.  The Langevin equation for the $x$
position of any monomer is $ x(t + \Delta t) =x (t) - \frac{\partial
V}{\partial x} \Delta t + \eta(t) \sqrt{ 2 k_BT \Delta t}$ where
$\eta(t)$ is the random noise chosen from a Gaussian distribution with
mean $0$ and standard deviation $1$.  The potential, $V$ includes all
the Lennard-Jones interactions along with the spring forces between
monomers.  For the spring forces we used the potential $V_{spring} =
kTb^{-1}\ell^*(r/(b))$ which is equivalent to a Gaussian chain
entropic spring potential $3kTr^2/(2b^2)$ at short extensions but
increases rapidly for extensions close to or beyond the bond length,
$b$. For practical purposes, we calculated $V_{spring}$ using a taylor
series expansion in $r$ to terms of order $r^{20}$. This gives a close
approximation to $V_{spring}$ over the range of $r$ in our simulation.
Other simulation parameters of importance were the time step, $\Delta
t=0.001$, temperature $k_BT=0.3$ and bond length $b = 2$. 

\begin{figure}
\onefigure[]{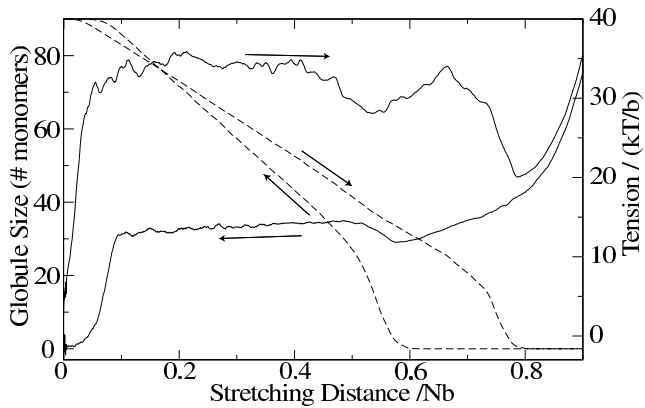}
\caption{ Simulation results for an average of 100 runs using $90$
monomers with a stretching speed of $0.001 b/t$.  The chain tension is
shown as a solid line and the number of monomers remaining in the ball
is shown as a dashed line.  Both stretching and relaxation processes
are shown with the direction of motion indicated with an arrow.
\label{simav}}
\end{figure}

An initial random walk configuration was allowed to collapse to a
globule and equilibrated for $10^6$ timesteps. It was then stretched
by applying a very small fixed movement $\Delta z$ on the $N$th
monomer at each timestep.  The $N$th monomer was otherwise free to
move in the $x$ and $y$ directions and the first monomer was fixed in
position.  Choice of stretching velocity in our simulations is a
delicate issue as it can affect the qualitative nature of the force
profiles.  We performed stretching simulations for a range of
stretching rates and found that the force profiles converged to a
limit for rates slower than $0.001 b/t$, where $b$ is the bondlength
and $t$ is the unit of time (t = 1000 time steps).  Such convergent
behaviour suggests that for stretching rates less than $0.001 b/t$ the
polymer is able to continuously relax to a local free energy minimium
as the extension is increased or decreased. To obtain average force
profiles we used a relatively fast rate of $0.001 b/t$ (see figure
\ref{simav}). A much slower rate of $1x10^{-4} b/t$ was used to obtain
results for a single run as close as practicable to local equilibrium
conditions (see figure \ref{fvsrsim}).  Nonetheless it is worth noting
that our results for the two stretching rates differed very little.

Our simulation results for the slower stretching rate (figure
\ref{fvsrsim}) confirm the existence of an unraveling transition
during pull out ( B to C in figure \ref{fvsrsim}) and a re raveling
transition during relaxation (D to E in figure \ref{fvsrsim}).
These transitions are clearly evident as a sharp change in the number
of monomers remaining in the ball, $N_c$ and also, to a lesser extent,
in the tension $T$.  There is marked hysteresis in the curves.  The
unraveling and re raveling transitions do not occur at the same
extension, in part due to the energy barrier between the states and
the finite time of the experiment.  In figure \ref{simav} we show
quantities averaged over 100 runs (figure \ref{simav}) for the
faster stretching speed.  As an aside, in these we can also see
clearly the transition to a tadpole at weak stretching predicted in
reference \cite{halperin_zhulina:1991}.

A recent experimental study on the stretching of DNA molecules using
optical tweezers obtained a very similar force profile to that in
figure \ref{fvsrsim} \cite{baumann_etal:2000}.  DNA is a charged,
stiff polymer and we would not expect perfect agreement with our
results, however similar physics seems to be at work.  As in our
results, Baumann {\it et al }\cite{baumann_etal:2000} observe a sudden
drop in the force, corresponding to the final unraveling. They also
observed an extension profile that is not monotonic but shows many
spikes in the region where the ball and chain configuration
exists. Our simulation results also showed such spikes, which are much
larger than the thermal noise. A direct hint of their origin is found
in the plots of the number of monomers remaining in the ball,
$N_c$. Each spike is associated with a sudden drop in $N_c$, so that
the spikes represent one or more monomers being pulled out of the
ball. During extension the monomers must of course be pulled out of
the ball more-or-less according to sequence along the chain. However
the random packing within a relaxed ball means that each monomer, as
well as its near neighbours along the chain is subjected to a rapidly
varying potential within the ball. This potential will have many
maxima and minima and to get over the maxima extra force must be
applied. Once this force is applied, one or more monomers pop out of
the ball and the force then drops suddenly. If this interpretation is
correct then during the gradual relaxation of the chain there should
be no spikes, since the monomers merely accrete to the outside of the
ball, and indeed this is what is seen. In order to investigate this
further we conducted simulations (not shown), where the chain was
rapidly and cyclically extended and relaxed. In these simulations,
once a chain had been extended and relaxed though one cycle, the
spikes on further extension cycles were much reduced. These spikes
contribute substantially to the hysteresis, since significant energy
is dissipated in the form of friction at the completion of each spike.
Similar spikes are seen in experiments\cite{baumann_etal:2000} and a
recent theoretical \cite{blumenfeld:2000} study has assumed
entanglements are the major cause.  Although this is certainly
possible for long chains, in our simulations, which had 90 monomers,
true entanglements must be fairly rare and in our case the rapidly
varying potential is probably the major cause.

Before we conclude, it is important to discuss another transition,
which was already mentioned in the paper by Halperin and Zhulina
\cite{halperin_zhulina:1991} . These and later authors using computer
simulation
\cite{wittkop_etal:1996,kreitmeier:2000,kreitmeier_etal:1999} assumed
that the chain was below, but close to the $\theta$ temperature so
that the solvent is only slightly poor. In this regime there is
substantial penetration of the solvent into the ball. A different kind
of transition can then occur when the number of monomers removed from
the the ball is such that the ball approaches the thermal ``blob"
size. In this case the ball has so few monomers that the collapsed
state is no longer the equilibrium state and it just evaporates.  We
would call this the vapourisation transition - it occurs because the
ball itself becomes unstable. This is in marked contrast to our
unraveling transition, for which the isolated ball is always stable.

%See fig.~\ref{f.1}, table~\ref{t.1} and eq.~(\ref{e.1}).
%See also~\cite{b.a,b.b}.

%\begin{figure}
%\onefigure{epl-template.eps}
%\caption{Figure caption.}
%\label{f.1}
%\end{figure}

\end{document}